\documentclass[runningheads]{llncs}
\usepackage{graphicx}
\usepackage{comment}
\usepackage{amsmath,amssymb} 
\usepackage{bm}
\usepackage{color}

\usepackage{multirow}
\usepackage{algorithm}
\usepackage{algpseudocode}

\usepackage{footnote}
\makesavenoteenv{tabular}
\makesavenoteenv{table}
\usepackage{bigstrut} 
\usepackage{rotating} 
\usepackage{makecell}

\usepackage{wrapfig,booktabs}
\usepackage{floatflt}

\usepackage{xr}
\makeatletter
\newcommand*{\addFileDependency}[1]{
  \typeout{(#1)}
  \@addtofilelist{#1}
  \IfFileExists{#1}{}{\typeout{No file #1.}}
}
\makeatother

\usepackage{footmisc}
\DefineFNsymbols{mySymbols}{{\ensuremath\ast}{\ensuremath\dagger}
{\ensuremath\ddagger}{\ensuremath\mathsection}{\ensuremath\mathparagraph}
{\ensuremath\parallel}{**}{\ensuremath{\dagger\dagger}}{\ensuremath{\ddagger\ddagger}}}
\setfnsymbol{mySymbols}

\begin{document}

\pagestyle{headings}
\mainmatter
\def\ECCVSubNumber{3891}  
\title{A Structure-Aware Relation Network for Thoracic Diseases Detection and Segmentation}

\titlerunning{SAR-Net}
\author{
	Jie Lian$^{1}$ \and
	Jingyu Liu$^{2}$ \and
	Shu Zhang$^{1}$ \and
	Kai Gao$^{1}$ \and
	Xiaoqing Liu$^{1}$ \and
	Dingwen Zhang$^{3}$\and
	Yizhou Yu\thanks{Yizhou Yu is corresponding author. This paper has been accepted by IEEE Transactions on Medical Imaging.}$^{4}$
}
\authorrunning{J. Lian et al.}
\institute{
$^1$Deepwise AI Lab\\
$^2$School of Electronics Engineering and Computer Science, University of Peking\\
$^3$School of Automation, University of Northwestern Polytechnical\\
$^4$Department of Computer Science, University of Hong Kong\\
\email{
lianjie@deepwise.com, jingyu.liu@pku.edu.cn, \{zhangshu@deepwise, gaokai@deepwise, liuxiaoqing@deepwise\}.com, zhangdingwen2006yyy@gmail.com, yizhouy@acm.org
}
}

\maketitle

\begin{abstract}
Instance level detection and segmentation of thoracic diseases or abnormalities are crucial for automatic diagnosis in chest X-ray images. Leveraging on constant structure and disease relations extracted from domain knowledge, we propose a structure-aware relation network (SAR-Net) extending Mask R-CNN. The SAR-Net consists of three relation modules: 1. the anatomical structure relation module encoding spatial relations between diseases and anatomical parts. 2. the contextual relation module aggregating clues based on query-key pair of disease RoI and lung fields. 3. the disease relation module propagating co-occurrence and causal relations into disease proposals.
Towards making a practical system, we also provide ChestX-Det, a chest X-Ray dataset with instance-level annotations (boxes and masks).
ChestX-Det is a subset of the public dataset NIH ChestX-ray14. It contains $\sim$ 3500 images of 13 common disease categories labeled by three board-certified radiologists. We evaluate our SAR-Net on it and another dataset DR-Private. Experimental results show that it can enhance the strong baseline of Mask R-CNN with significant improvements. The ChestX-Det is released at \url{https://github.com/Deepwise-AILab/ChestX-Det-Dataset}.
\end{abstract}

\section{Introduction}
Chest X-ray scan is a routine examination for thoracic diseases in hospitals. With domain expertise, radiologists can identify and localize abnormalities or diseases for further diagnosis. To reduce the burden on radiologists, computer-aided diagnosis is put into increasing efforts in recent years. With the success of deep Convolutional Neural Network (CNN) on natural images, applications like classification, detection and segmentation in medical images are also overwhelmingly benefited. In this paper, we aim to detect and segment thoracic abnormalities at the instance level based on Mask R-CNN\cite{MaskRCNN}. Combining domain knowledge extracted from chest X-ray studies, we extend Mask R-CNN, a successful instance-segmentation framework, with our relation modules. To clarify in the following, we use the term ``abnormality'' and ``disease'' alternatively as ``object'' in natural images.

Our relation modules are motivated from three types of relations: \textbf{1. Spatial relations between diseases and thoracical anatomical structures.} Diseases often have location priors. Encoding spatial relations and constraints can help obtain more accurate locations.
\textbf{2. Contextual relations between abnormalities and observation in lung fields.} Contextual clues are always useful for radiologists. One typical example is contralateral examination, e.g., over-exposured xrays have similar appearance with lung consolidation. By checking contralateral appearance, computers can mimic radiologists to exclude this type of alarms. \textbf{3. Categorical dependent relations among diseases.} It is common knowledge that one disease can cause another disease. Also, a complex disease might be caused by a combination of factors, resulting in various abnormalities. Therefore abnormalities can co-exist in one x-ray image.

\begin{figure}[ht]
	\begin{center}
		\includegraphics[scale=0.5]{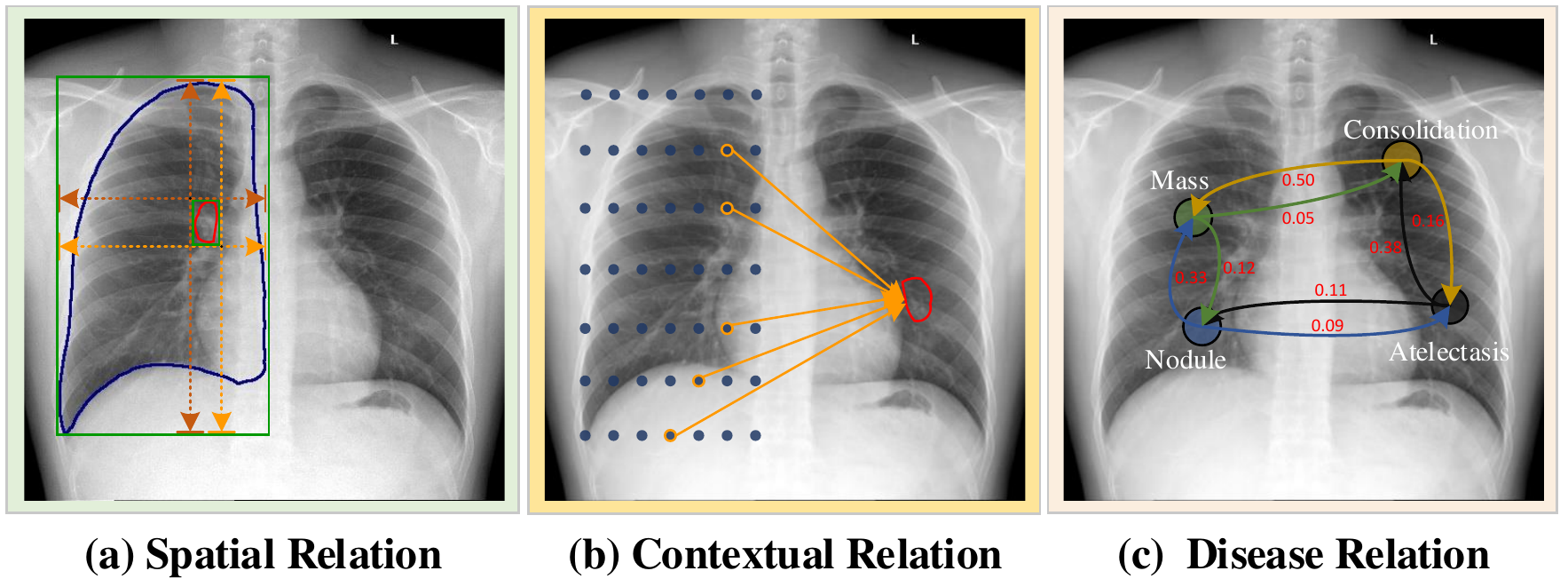}
	\end{center}
	\caption{
		Three types of relations in SAR-Net. (a) Spatial relation with anatomical parts. Firstly, a pre-trained segmentation model is adopted to obtain the anatomical parts (Lung Fields, Scapulae and Heart).Then for each disease RoI, using coordinate difference to quantify its spatial relation with each anatomical part. (b) Contextual relation bounded in lung fields. After the grid-shaped features of both lung fields are obtained, the contextual relation module models contextual relations between each disease RoI and observations in the lung fields. (c) Disease relation of categorical dependence. Message passing via relation graph to enhance the features of each disease RoI. The relation graph contains co-occurrence and causal relations among diseases and message is the semantic embedding of each category disease.}
	\label{fig:introduction}
\end{figure}

To this end, we propose a structure-aware relation network which consists of three modules: spatial relation module, contextual relation module and disease relation module.
We first extract the anatomical structures via a pre-trained semantic part segmentation model. Instead of segmentation masks, we adopt only bounding boxes of anatomical parts for further computation. 
For the spatial relation module, we encode the spatial relations between disease proposals and anatomical parts. For the contextual relation module, we extract lung-field features as the context, then adopt the query-key attention mechanism to learn the attention weights in the lung fields. Finally contextual features are weighted aggregated. For the disease relation module, we first build a relation graph among diseases based on the co-occurrence frequency between each pair of diseases. Then we propagate information via the adjacent matrix among diseases. 
Figure \ref{fig:introduction} illustrates three types of relations.
The outputs of all three modules are encoded as feature vectors for each disease RoI proposal. We concatenate them with original RoI features after RoI-pooling. Both the box head and mask head are trained from scratch to obtain more accurate detection and segmentation results.

\begin{figure*}[t]
	\begin{center}
		\includegraphics[scale=0.145]{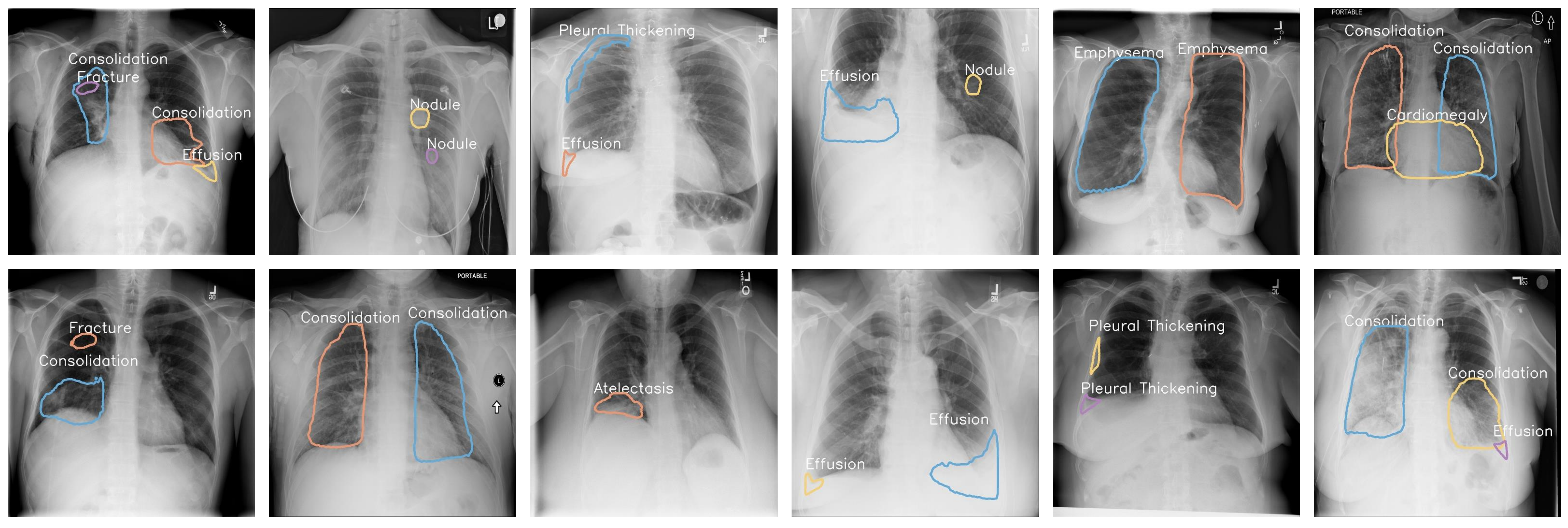}
	\end{center}
	\caption{Some instance segmentation results on ChestX-Det. To retain disease appearance, we only show their contours.}
	\label{fig:seg}
\end{figure*}

Above all, the contribution of our work are three folds: 1. To push forward the research on fully supervised instance-level detection and segmentation on chest X-Rays, we provide a new benchmark called ChestX-det, including instance-level annotations of 13 categories of disease/abnormality of $\sim$ 3,500 images from the public dataset NIH ChestX-ray14. 2. We provide a strong baseline of Mask R-CNN that can be compared for further research. 3. We propose SAR-Net modeling three types of relations. It can be embedded into general detection frameworks and enhance the baseline with significant improvements.

\section{Related Works}
\subsection{Automatic Chest X-ray Analysis}
Most existing works on chest X-rays focus on disease classification and weakly supervised localization\cite{ChestX8,Chexnet,guan2018diagnose,pesce2017learning,ypsilantis2017learning,yan2018weakly}, due to the fact that most current public datasets contain only class labels or very limited amounts of box annotations. CheXpert\cite{Chexpert} proposed by Irvin et al. contains 224,316 frontal and lateral view chest X-ray images. In training set, the labels (1/0/uncertain of the presence of 14 classes of observations) are extracted from radiology reports by a carefully designed labeler. In validation and test datasets, the labels are labeled by radiologists. Earlier, NIH Chest X-ray 14\cite{ChestX8} proposed by Wang et al. contains 112,120 front-view images of 14 disease categories, among which there are 880 images of 8 categories containing box annotations. 
Lately, Nguyen et al. proposed VinDr-CXR \cite{VinDr} which contains 18,000 images that were manually annotated with 22 classes of rectangles surrounding abnormalities and 6 global labels of suspected diseases. 
There also exist some datasets that focus on a single disease, such as the Pneumonia detection dataset
$\footnote{https://www.kaggle.com/c/rsna-pneumonia-detection-challenge}$, Tuberculosis detection dataset \cite{tb_detection} and Pneumothorax segmentation dataset $\footnote{https://www.kaggle.com/c/siim-acr-pneumothorax-segmentation}$, etc.
CAM\cite{CAM}, grad-CAM\cite{Grad-CAM}, and attention models\cite{pesce2017learning,ypsilantis2017learning} are the most widely used tools to highlight rough locations of diseases. To use box annotations, Li et al. \cite{Lizhe} propose an end-to-end fully convolutional neural network to classify and localize abnormalities. The problem is addressed in an approach like weakly supervised segmentation at the coarse level. Based on \cite{Lizhe}, Liu et al\cite{LiuJY} propose a contrast induced attention network to predict the location of diseases, and an alignment module to align positive and negative images in the first place.
Our work is different that we are the first to detect chest X-ray diseases with fully supervised annotations.
Chen et al. \cite{Chen} present a deep hierarchical multi-label classification approach for Chest X-ray diagnosis. Inspired by the idea of taxonomy in \cite{Chen}, we group 13 categories into three parent classes (Table \ref{tab:datasets}) and evaluate the effect of each relation module based on the class hierarchy. More details are presented in the 4.6 subsection.

\subsection{Object Detection/Instance Segmentation}
Modern object detection frameworks include two lines of approaches. The first line is two-stage detectors of Fast R-CNN\cite{FastRCNN}, Faster R-CNN\cite{FasterRCNN}, FPN\cite{FPN}, etc. 
By using a top-down pathway and the lateral connections, FPN can enrich the semantic information of shallow layers while maintaining their high resolution. Utilizing multi-level features to improve performance is crucial in detection tasks.
The second line is one-stage detectors of YOLO\cite{YOLO}, SSD\cite{SSD} and recent anchor-free detectors\cite{CornerNet,FCOS,FSAF,CenternetTriplets,CenterPoints}. We choose FPN as our backbone framework since disease regions in chest X-rays range from 20$\times$ 20 pixels (e.g. nodule, calfication) to 1000$\times$ 1000 pixels (e.g. emphysema). FPN is known for extracting RoI features from shallow to deep layers, capturing characteristics from small to large objects. Mask R-CNN\cite{MaskRCNN} is a successful instance segmentation framework extending FPN. The mask head operating on RoIs can output class-specific masks. Since disease regions are always of various shapes, we use mask R-CNN to visualize them for more accurate diagnosis. However, one difference for chest X-ray images is the constant anatomical structure, which is not available in natural images. Furthermore, general object detectors like FPN and Mask R-CNN fail to capture object-wise and class-wise relations, let alone object-structure relations.
There exists works exploring relations for general object detection. In \cite{RelationNet}, appearance and geometry relations within objects are modeled inspired by attention modules\cite{Attention} in NLP. In \cite{ReasonRCNN}, object features are enhanced via attending different semantic concepts and propagating information through a common sense knowledge graph. In \cite{SpatialGraph}, a sparse graph is built based on semantic and spatial relations among objects. Our work is closely related to them while we focus on the relations between diseases and structures.

\begin{figure*}[t]
	\begin{center}
		\includegraphics[scale=0.385]{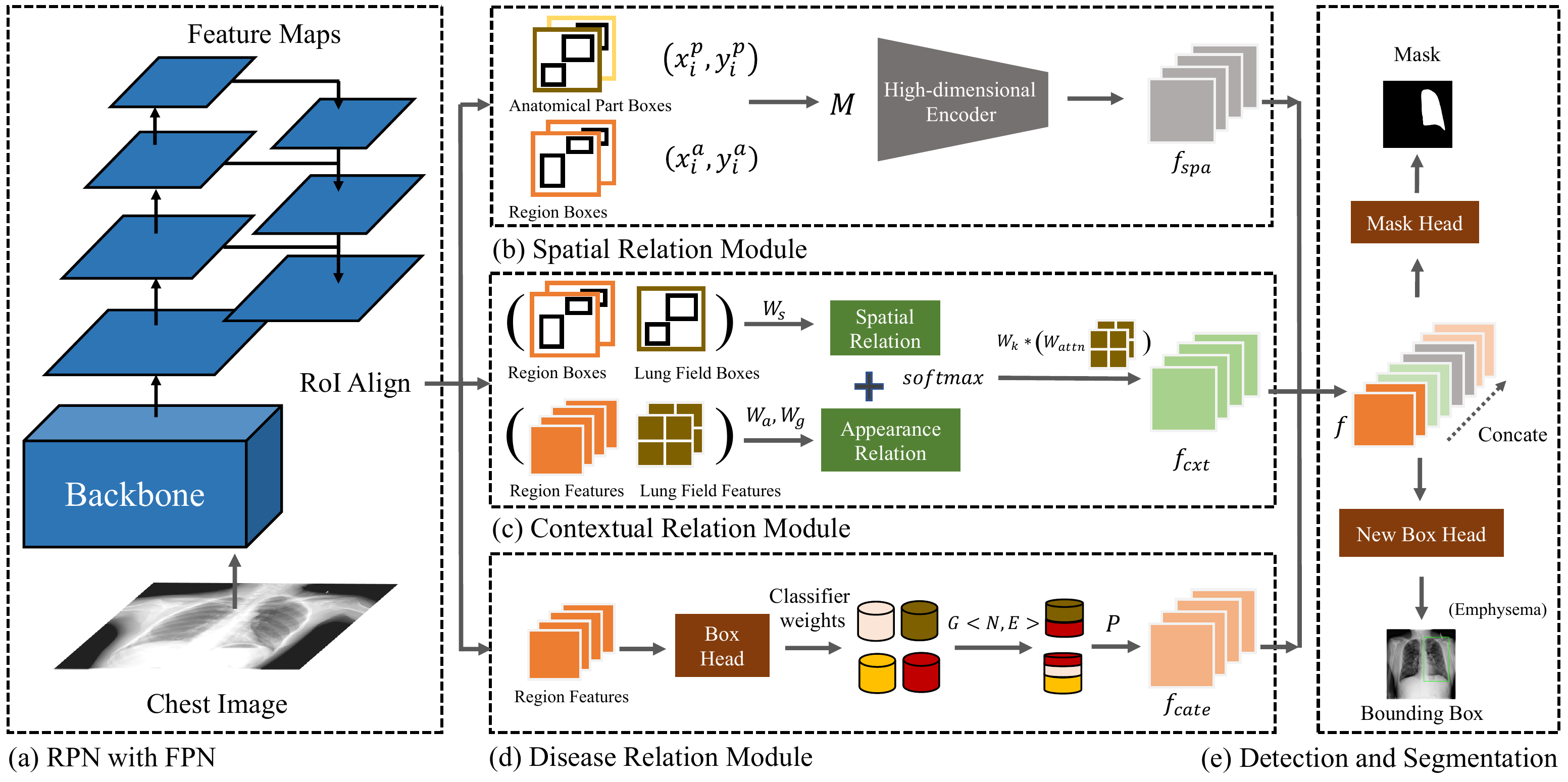}
	\end{center}
	\caption{Framework of our proposed SAR-Net. (a)Features from RoIs of anatomical parts are extracted from P4 of FPN. Disease region proposals are generated by RPN from corresponding layers. (b)Spatial relations between disease proposals and anatomical parts are encoded to $f_{spa}$. (c)Contextual relations between disease proposals and contexts in lung fields are encoded to $f_{cxt}$. Attention weights are computed based on spatial and appearance relations. (d)Co-occurrence and causal relations within categories are propagated to formulate $f_{cate}$. (e)Enhanced RoI features via concatenating $(f:f_{cxt}:f_{spa}:f_{cate})$ are input to box head and mask head to generate bounding boxes and masks.}
	\label{fig:framework}
\end{figure*}

\section{Method}
In this section, we present our SAR-Net (Structure-Aware Relation Network).
The SAR-Net (Figure \ref{fig:framework}) consists of three relation modules: anatomical structure relation module, contextual relation module and disease relation module.
For $N_r$ region proposals with their feature $f(N_r)\in\mathbb{R}^{N_r \times D}$, our aim is to enhance each region proposal feature $f$ by concatenating $f_{spa}$, $f_{cxt}$ and $f_{cate}$ computed from the three modules.

As shown in  Figure \ref{fig:framework}, all the five parts are integrated into the whole framework, and are trained end-to-end. To be more specific, additional features from three relation modules are online. During the iterative training of RPN with FPN, locations (Region Boxes) and visual appearance (Region Features, Lung Field Features) keep changing online.
Hence, the spatial, contextual and categorical relations ($f_{spa}$, $f_{cxt}$ and $f_{cate}$) are also changing online.
Architectures of three relation modules will be detailed in following subsections respectively.

\subsection{Anatomical Structure Relation Module}
Some diseases or abnormalities are highly correlated with specific parts or organs in the body. For instance, cardiomegaly is a medical condition in which the heart is enlarged; atelectasis is a complete or partial collapse of the entire lung or area (lobe) of the lung.

Based on the above observation, we propose an anatomical structure relation module to encode spatial relations between disease and anatomical parts.
To accomplish the task, we first adopt a pre-trained segmentation model $\footnote{https://github.com/Deepwise-AILab/ChestX-Det-Dataset/tree/main/pre-trained\_PSPNet}$ to obtain anatomical parts.
Then we choose five key parts of left lung, right lung, left scapula, right scapula and heart, as shown in Figure \ref{fig:parts}.
The segmentation model is trained on 1000 chest images labeled with 5 parts from external data, and is adopted to generate anatomical parts for each image in our datasets. Then for each disease RoI, we use coordinate difference to quantify its spatial relation with each anatomical part. 

\begin{floatingfigure}[r]{0.46\textwidth}
	\centering
	\includegraphics[width=0.46\textwidth]{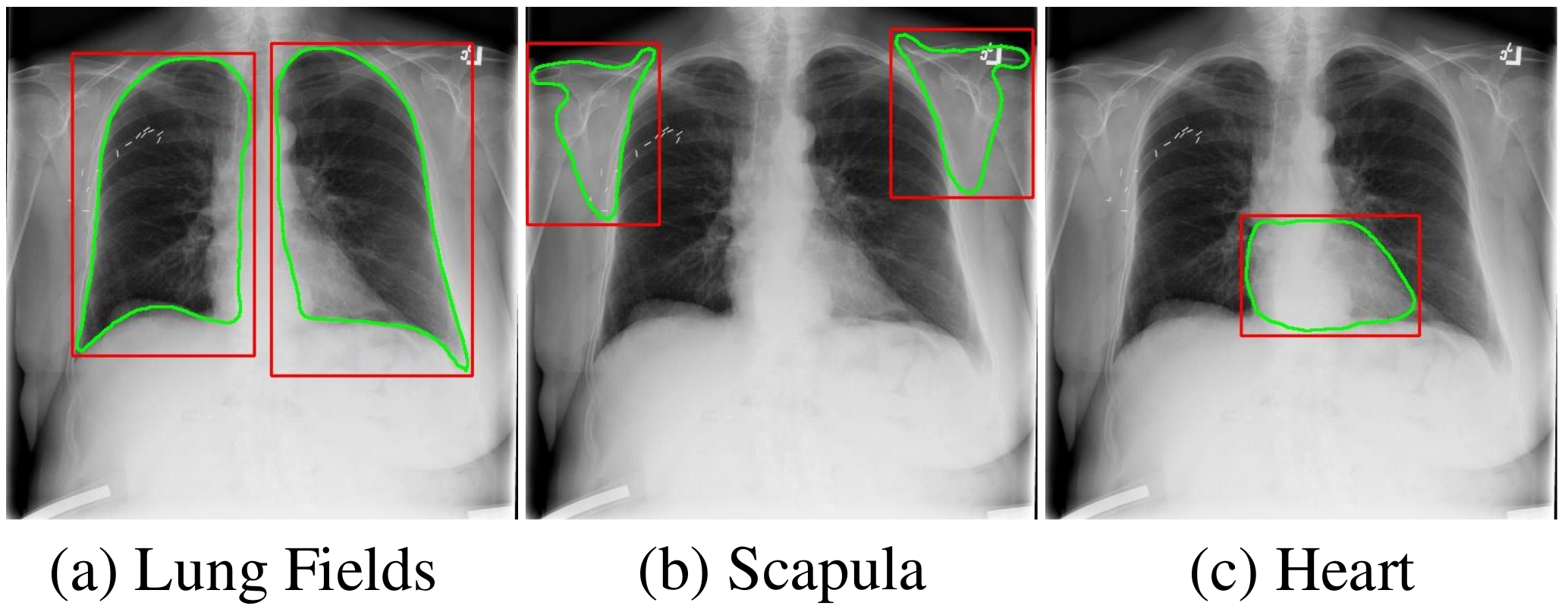}
	\caption{Segmented anatomical parts via pretrained PSP-Net \cite{PSPNet}. We use five parts (left\&right lung field, left\&right scapulae, and heart) in the spatial relation module. We use left\&right lung field to bound contextual areas.}
	\label{fig:parts}
\end{floatingfigure}

\begin{equation}
\label{E6}
\begin{split}
M_p =& (\frac{x_1^a-x_1^p}{w_l}, \frac{y_1^a-y_1^p}{h_l}, \frac{x_1^a-x_2^p}{w_l}, \frac{y_1^a-y_2^p}{h_l}, \\
& \frac{x_2^a-x_1^p}{w_l}, \frac{y_2^a-y_1^p}{h_l}, \frac{x_2^a-x_2^p}{w_l}, \frac{y_2^a-y_2^p}{h_l})
\end{split}
\end{equation}
where $\left\{ (x_{i}^{a}, y_{i}^{a}) \right\}_{i=1}^{2}$ and $\left\{ (x_{i}^{p}, y_{i}^{p}) \right\}_{i=1}^{2}$ are up-left and bottom-right vertex coordinates of disease RoI and anatomical part respectively,
and $w_{l}$ and $h_{l}$ are width and height of the bounding box covering left and right lung.
Relating all five parts, $M=Concat\left\{ (M_p)_{p=1}^5 \right\}$ is a 40-dimensional vector.
Inspired by \cite{Attention}, we embed $M$ into a high-dimensional space, as
\begin{equation}
\label{E7}
\begin{split}
f_{spa} = Concat[sin(M / 1000^{j/d_e}), \qquad \qquad\\ cos(M / 1000^{j/d_{e}})], \quad j \in [0, d_e-1]
\end{split}
\end{equation}
where sine and cosine functions of different wavelengths are computed. We can change the dimension of $f_{spa}$ by setting different $d_e$. Finally, the spatial relation between all disease proposals and anatomical parts are recorded in position embedding $f_{spa}(N_r) \in \mathbb{R}^{N_r\times D_{spa}}$, where $D_{spa} = 80 *d_e$.

\subsection{Contextual Relation Module}
Contextual cues are very useful for radiologists to reference. For instance, by checking on the symmetric area on the contralateral side, radiologists can decide if the abnormality is a nodule or simply papilla. In our work, we focus on diseases in lung fields. Therefore we bound contextual area into the anatomical parts of left and right lung. Here we customize the attention mechanism having been proven successful in natural language processing and natural image recognition. Similarly, for a query of a particular disease RoI, a set of key contents in the lung fields are aggregated according to attention learned on disease-context relations. The relations involve both spatial and feature compatibility.

Specifically, we use $M_l$ and $M_r$ to denote the bounding box of left lung and right lung respectively. 
We adopt the same operation on $M_l$ and $M_r$ to obtain their grid-shaped feature $f_l$ and $f_r$, of size $N_d^2 \times D_M$. For each disease RoI feature $f_{a_i}$ (after 2 fc layers) and each grid feature $f_{g_j}$, we compute their compatibility as
\begin{equation}
\label{E9}
\mathcal{F}(a_i,g_j) = \left \langle W_a f_{a_i}, W_g f_{g_j} \right \rangle
\end{equation}
where $W_a$ and $W_g$ are both weight matrices, and $W_a f_{a}$ and $W_g f_{g}$ share the same dimension.

Similar with Eq.\eqref{E6} and Eq.\eqref{E7}, we can obtain the spatial relation between abnormality $a_i$ and grid $g_j$. 
\begin{equation}
\label{E8}
\mathcal{P}(a_i,g_j) = ReLU(W_s {s(a_i,g_j)})
\end{equation}
where $s(a,g)$ is $(\frac{x_1^a-x^g}{w_l}, \frac{y_1^a-y^g}{h_l}, \frac{x_2^a-x^g}{w_l}, \frac{y_2^a-y^g}{h_l})^T$ $\footnote{$(x_i^a, y_i^a)$ are coordinates of up-left and bottom-right vertexes of abnormality RoI. $(x^g, y^g)$ is the center coordinate of a grid.}$, and $W_s\in\mathbb{R}^{1\times 4}$ is a learnable weight matrix.

Then, the spatial relation $\mathcal{P}(a_i,g_j)$ and appearance relation $\mathcal{F}(a_i,g_j)$ are collected across all grids and summed into a soft-max function to get attention weights as 
\begin{equation}
\label{E10}
\omega_{attn}^j = \frac {exp \big( log(\mathcal{P}(a_i,g_j)) + \mathcal{F}(a_i,g_j) \big)}{\sum_k exp \big( log(\mathcal{P}(a_i,g_k)) + \mathcal{F}(a_i,g_k) \big)}
\end{equation}

Finally, grid-wise contextual features from lung fields are aggregated according to attention learned above.
\begin{equation}
\label{E11}
f_{cxt} = W_{k} * (\sum_{j=1}^{2\times N_d^2} \omega_{attn}^j\left[f_l, f_r\right]_j)
\end{equation}
where $W_{k}$ denotes the $1\times 1$ convolution kernel weights. The enhanced features of all region proposals are $f_{cxt}(N_r) \in \mathbb{R}^{N_{r} \times D_{cxt}}$. 
Note that [,] means the concatenation operation.

\subsection{Disease Relation Module}
\label{section:DRM}
Diseases or abnormalities in chest X-Rays are highly correlated. For instance, pulmonary tuberculosis is a complex disease which might contain nodules, fibrosis and consolidation simultaneously. There also exists causal relations among abnormalities. For instance, rib fracture is likely to cause pneumothorax.
To this end, first we need to build a relation graph containing co-occurrence and causal relations among diseases.
Then messages from semantic concepts of diseases are propagated via the relation graph.
Finally RoI features aggregate messages to form representation $f_{cate}$.
These three steps are detailed in the following.

\textbf{Relation Graph Construction.}
To build the relation graph, we count co-occurrence frequency in the data set of each pair of categories and compute their conditional probability as
\begin{equation}
\label{E_prob}
P(A|B) = \frac{\# (A, B)}{\# (B)}
\end{equation}
where $\# (B)$ denotes number of samples disease B exists, and $\# (A, B)$ denotes number of samples disease A and B co-exist.
Then we build a directed graph ${G} = <N, E>$ as the disease relation graph.
$N$ are disease category nodes and each edge $e_{i,j} \in E$ encodes relations:
\begin{equation}
\label{E_edge}
e_{A \to B} = P(B|A),  \quad e_{B \to A} = P(A|B)
\end{equation}

\textbf{Message Passing via Relation Graph.}
The relation graph is built based on statistics collected from thousands of X-rays and reflects prior knowledge.
However, for a specific sample, it only reflects the condition of a patient in a limited period.
Therefore, we propose to use a global attention to force message passing on potential diseases.

First, we input the whole X-ray image into the same convolutional layers of SAR-Net to get the image feature $F \in \mathbb{R}^{H \times W \times D^{'}}$.
Then a global mean-average-pooling operation is applied to squeeze $F$ into $F_{s} \in \mathbb{R}^{D^{'}}$.
Finally, we use a fully-connected layer to get categories score $\beta \in \mathbb{R}^C$, where $C$ is the number of categories.
The global binary cross-entropy loss for each category is defined as 
\begin{equation}
\label{E1}
\textit{L(c)} = \sum_{i \in C} -y_{i}log(p_{i}) -  \sum_{i \in C} (1-y_{i})log(1-p_{i})
\end{equation}
where $i$ is the index of categories, $y_{i}$ denotes the target label of the category and $p_{i}$ denotes the predicted probability.

Inspired by some few/zero-shot works \cite{Clsweight1,Clsweight2,Clsweight3}, we use parameter weights of the classifier branch in box head to represent the semantic embedding of each category. Formally, the semantic embedding is defined as $W_{emb} \in \mathbb{R}^{C \times D}$, where $C$ and $D$ are number of categories and weight dimensions respectively.
By transmitting causal or co-occurrence relations through the sparse relation graph, we can obtain the $i$-th category embedding as 
\begin{equation}
\label{E}
z_{i} = \sum_{j \in C} \beta_{j}e_{j \to i}w_{j} 
\end{equation}

\begin{figure*}[t]
	\begin{center}
		\includegraphics[scale=0.335]{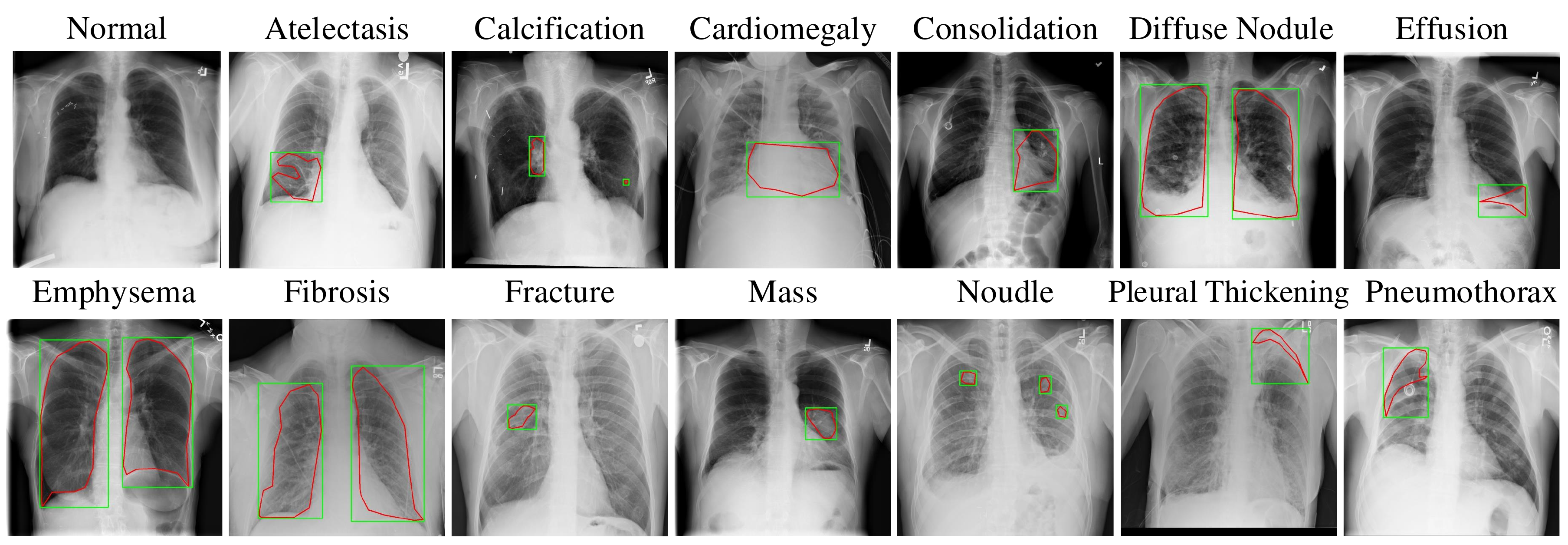}
	\end{center}
	\caption{Examples of annotated diseases of 13 categories in ChestX-Det. The first example is a ``normal'' image without these specific disease.}
	\label{fig:disease}
\end{figure*}

\textbf{Mapping Disease Relation to Regions.}
Since our goal is to enhance the original region features, we need to map the global embedded semantics from $C$ categories to $N_{r}$ regions. In our work, we choose the classification probability of each RoI $P \in \mathbb{R}^{N_{r} \times C}$ as the mapping bridge. The semantic embedding of $k$-th region is
\begin{equation}
\label{E3}
r_{k} = \sum_{i \in C} p_{ki}z_{i}
\end{equation}
where $p_{ki}$ denotes the probability for the $k$-th region towards category $i$. 

Finally, $r_k$ is linearly transformed to reduce dimension
\begin{equation}
\label{E5}
f_{cate} = W_t r_k
\end{equation}
where $W_{t}$ is the weight of a fully-connected layer. The enhanced features of all regions are $f_{cate}(N_r) \in \mathbb{R}^{N_{r} \times D_{cate}}$.
Now, we obtain the embedded feature of spatial relation $f_{spa}$, contextual relation $f_{cxt}$ and category relation $f_{cate}$. The final enhanced feature of each disease RoI is $f' = [f:f_{spa}:f_{cxt}:f_{cate}], f'(N_r) \in \mathbb{R}^{N_{r} \times (D+D_{spa}+D_{cxt}+D_{cate})}$.

\section{Experiments}
\subsection{Dataset and Evaluation}
\textbf{Datasets:}
\textbf{ChestX-Det} is a subset with box annotations of NIH ChestX-14. NIH ChestX-14 contains 112,120 front-view images, among which there are 880 images with box annotations.
To make fully supervised learning feasible, we select 3,575 images and invite three board-certified radiologists to annotate them with 13 common categories of diseases or abnormalities.
For annotation, we split three board certified radiologists into two roles.
\textbf{Commitee:} Every chest X-ray is annotated by the first two radiologists. The two radiologists are mutually blind to the annotation. \textbf{Judge:} The third and the most experienced radiologist with 15+ years of experience select annotations from those provided by the first two radiologists, and can add annotations if there are more.
Figure \ref{fig:disease} shows examples of the annotated 13 categories and a ``normal'' sample without abnormalities.
We use 3025 images for training and 553 images for testing. 10\% samples from the training set are used for validation. 

\begin{floatingtable}[r]
	{
	\begin{scriptsize}
		\resizebox{80mm}{28mm}{
		\begin{tabular}{c|c|c|c}
			\hline 	\hline
			\multirow{2}[3]{*}{Parent Classes} & \multirow{2}[3]{*}{Abnormalities} & ChestX-Det & DR-Private \bigstrut[b]\\
			\cline{3-4}          &       & Train\&Val / Test & Train\&Val / Test \bigstrut\\
			\hline
			\multirow{9}[2]{*}{LUNG} & Atelectasis & 289 / 51 & 265 / 41 \bigstrut[t]\\
			& Calcification & 281 / 67 & 574 / 71 \\
			& Consolidation & 2110 / 453 & 1904 / 291 \\
			& Diffusive Nodule & 195 / 51 & 181 / 33 \\
			& Emphysema & 232 / 66 & 344 / 49 \\
			& Fibrosis & 619 / 120 & 780 / 102 \\
			& Fracture & 547 / 115 & 4407 / 643 \\
			& Mass  & 133 / 34 & 319 / 43 \\
			& Nodule & 848 / 182 & 720 / 156 \bigstrut[b]\\
			\hline
			\multirow{3}[2]{*}{PLEURA} & Effusion & 1734 / 379 & 1413 / 226 \bigstrut[t]\\
			& Pleural Thickening & 526 / 105 & 512 / 82 \\
			& Pneumothorax & 169 / 42 & 1095 / 130 \bigstrut[b]\\
			\hline
			MEDIASTINUM & Cardiomegaly & 223 / 70 & 653 / 140 \bigstrut\\
			\hline
			\hline
		\end{tabular}}
	\end{scriptsize}
}
\caption{The statistics of abnormalities.}
\centering
\label{tab:datasets}%
\end{floatingtable}%

\textbf{DR-private} is a private dataset by collecting 6,629 chest X-ray images from multiple Chinese hospitals. The annotation process is the same with ChestX-Det.
As for DR-private, we use 5800 images for training and 829 images for testing. 10\% samples from the training set are used for validating.
Table \ref{tab:datasets} shows the instance count of each disease in both datasets. For both instance segmentation and anatomical part segmentation, all data are selected to cover the appearance range as wide as possible. We include normal and abnormal samples in different angles, whiteness, and scanning conditions. For white-out lungs in part segmentation, we ask radiologists to annotate their original contours as far as they can.

\textbf{Part Segmentation Performance:}
We evaluate the part segmentation model on 229 test samples, the mean IOU (Mean Intersection-Over-Union) is 86.85\%. We only use the bounding boxes of segmentation for further computation. 
The accuracy of segmentation does not seriously affect the bounding boxes.
The segmentation results of PSP-Net can well serve the SAR-Net. 

\textbf{Evaluation:}
We borrow the metric of bounding box AP (AP$^{bb}$) \cite{AP} used in general object detection. Considering that disease/abnormality regions lack clear boundary as general objects, we adopt AP$^{bb}_{50}$ as the main evaluation metric. AP$^{bb}_{25}$ and AP$^{bb}_{75}$ are used for reference. We believe that AP$^{bb}_{75}$ better reflects localization performance and AP$^{bb}_{25}$ better reflects classification performance. AP$^{bb}_{50}$ is the comprehensive performance index for classification and localization. For more practical usage, we also present instance-level recall at fixed FP(false positive) per image for more direct reference. 
In addition, we use mask AP (AP$^{mask}$) to evaluate the performance of instance segmentation. AP$^{mask}_{50}$ and AP$^{mask}_{75}$ are also provided for reference.

\subsection{Experiment Setup}
We implement SAR-Net stacked on six backbones: ResNet-50-C4\cite{ResNet} in Faster R-CNN\cite{FasterRCNN}, Mask R-CNN\cite{MaskRCNN} and Cascade R-CNN\cite{cascadeRCNN} respectively; ResNet-50-FPN\cite{FPN} in Faster R-CNN and Mask R-CNN respectively;
ResNet-50-FPN+DCN\cite{DCN} in Mask R-CNN. 	
The core idea of DCN is to capture position-based semantic context to improve the capacity of feature representation. It adds 2D offsets to the regular grid sampling locations and defines a new grid to select input values.

\begin{floatingtable}[r]
		{
		\begin{scriptsize}
			\resizebox{60mm}{33mm}{
				\begin{tabular}{c|c|c}
					\hline
					Modules & Parameters & Dimensions \bigstrut\\
					\hline
					\hline
					\multirow{4}[2]{*}{Spatial Relation} & $M_{p}$  & 8 \bigstrut[t]\\
					& $M$     & 40 \\
					& $d_e = 8$ & / \\
					& $f_{spa}$ & 640 \bigstrut[b]\\
					\hline
					\multirow{7}[2]{*} {Contextual Relation} & $f_{l,r}$ & $7^2\times256$  \bigstrut[t]\\
					& $W_a$  & $1024\times1024$ \\
					& $f_a$  & 1024 \\
					& $W_g$  & $256 \times 1024$ \\
					& $f_g$   & 256 \\
					& $W_k$  & $256 \times 1024$ \\
					& $f_{cxt}$ & 1024 \bigstrut[b]\\
					\hline
					\multirow{4}[2]{*}{Disease Relation} & $F_{s}$  & 256 \bigstrut[t]\\
					& $W_{emb}$ & $13 \times 1024$ \\
					& $W_{t}$  & $1024 \times 256$ \\
					& $f_{cate}$ & 256 \bigstrut[b]\\
					\hline
					Region Proposal Feature & $f$     & 1024 \bigstrut\\
					\hline
			\end{tabular}}%
		\end{scriptsize}}
			\caption{Parameter dimensions in three modules with FPN+DCN.}
	\label{parameters}%
\end{floatingtable}%

All experiments are implemented using Pytorch on 4 TITAN-V GPUs. ResNet-50 is pretrained on ImageNet \cite{ImageNet}. For all training, we apply stochastic gradient descent (SGD) with a weight decay of 0.0001 and momentum of 0.9 to optimize all models. The first conv layers of FPN and C4 are frozen. We train 50 epochs with image batch-size of 2 on each GPU. The learning rate starts at 0.01, and reduce by a factor of 10 after 20 and 40 epochs.

During training, we adopt random flipping and multi-scale sampling (shorter side=$\left\{ 800 \sim 1400 \right\}$) for all images. At testing stage, the shorter side of image is fixed at 1200. The total number of proposed regions $N_r$ after NMS is 512.
All the other hyper parameters and loss functions follow traditional Mask R-CNN.
As shown in Table\ref{parameters}, we clarify parameter dimensions in three modules with FPN-DCN. The extracted features $f_{cate}$,  $f_{spa}$ and $f_{cxt}$ from three relation modules are all 1D. All the weight parameters are initialized using gaussian prior.

\subsection{Comparison with the baseline model}
Table\ref{tab1} shows detection results comparison of our SAR-Net and baseline of various models on both datasets.
We can see that SAR-Net consistently outperforms baseline with any configuration of models, evaluation metrics and datasets.
Improvements on all models demonstrate that SAR-Net can be embedded into general two-stage object detection frameworks.

\begin{table*}[t]
	\centering
	\caption{Main detection results of test datasets on ChestX-Det and DR-private.}
	\resizebox{\textwidth}{!}{
		\begin{tabular}{c|c|c|c|c|c|c|c|c|c}
			\hline
			Model & Method & \%    & AP$^{bb}_{25}$  & AP$^{bb}_{50}$  & AP$^{bb}_{75}$  & \%    & AP$^{bb}_{25}$  & AP$^{bb}_{50}$  & AP$^{bb}_{75}$ \bigstrut\\
			\hline
			\hline
			\multirow{2}[2]{*}{Faster R-CNN(C4)} & Baseline & \multirow{12}[12]{*}{\begin{sideways}Chest X-Det\end{sideways}} & 50.5  & 36.5  & 10.4  & \multirow{12}[12]{*}{\begin{sideways}DR-Private\end{sideways}} & 42.8  & 33.6  & 11.9 \bigstrut[t]\\
			& SAR-Net &       & \textbf{$\bm{51.1}$} & \textbf{$\bm{38.2}$} & \textbf{$\bm{12.2}$} &       & \textbf{$\bm{46.3}$} & $\bm{36.6}$  & $\bm{12.9}$  \bigstrut[b]\\
			\cline{1-2}\cline{4-6}\cline{8-10}    \multirow{2}[2]{*}{Faster R-CNN(FPN)} & Baseline &       & 52    & 41.7  & 12.3  &       & 48.5  & 38.7  & 14.3 \bigstrut[t]\\
			& SAR-Net &       & $\bm{53.7}$  & $\bm{43.7}$  &$\bm{14.4}$  &       & $\bm{49.2}$ & $\bm{40.0}$ & $\bm{16.5}$  \bigstrut[b]\\
			\cline{1-2}\cline{4-6}\cline{8-10}    \multirow{2}[2]{*}{Cascade R-CNN(C4)} & Baseline &       &  48.6     &  39.0     &   15.1    &       & 41.9      & 32.5      & 14.0 \bigstrut[t]\\
			& SAR-Net &       & $\bm{51.2}$     & $\bm{41.7}$    & $\bm{16.4}$      &       &$\bm{43.1}$      & $\bm{34.6}$      &$\bm{15.4}$   \bigstrut[b]\\
			\cline{1-2}\cline{4-6}\cline{8-10}    \multirow{2}[2]{*}{Mask R-CNN(C4)} & Baseline &       & 48.6  & 37.9  & 10.9  &       & 42.6  & 32.3  & 13.3 \bigstrut[t]\\
			& SAR-Net &       &$\bm{51.4}$  & $\bm{40.8}$  & $\bm{13.4}$  &       & $\bm{45.3}$ &$\bm{35.7}$   &$\bm{15.3}$   \bigstrut[b]\\
			\cline{1-2}\cline{4-6}\cline{8-10}    \multirow{2}[2]{*}{Mask R-CNN(FPN)} & Baseline &       & 52.8  & 42.6  & 15.2  &       & 48.5  & 39.7  & 15.7 \bigstrut[t]\\
			& SAR-Net &       &$\bm{55.2}$  & $\bm{45.3}$ & $\bm{16.7}$  &       & $\bm{49.9}$  &$\bm{41.6}$   & $\bm{17.5}$  \bigstrut[b]\\
			\cline{1-2}\cline{4-6}\cline{8-10}    \multicolumn{1}{c|}{\multirow{2}[2]{*}{Mask R-CNN(FPN+DCN)}} & Baseline &       & 54.8  & 45.0  & 16.1  &       & 48.7  & 40.1  & 16.1 \bigstrut[t]\\
			& SAR-Net &       & $\bm{57.0}$  & $\bm{47.7}$ & $\bm{19.0}$  &       &$\bm{49.6}$   & $\bm{42.1}$  &$\bm{17.4}$  \bigstrut[b]\\
			\hline
	\end{tabular}}%
	\label{tab1}%
\end{table*}%

Stable gains are achieved at AP$^{bb}_{50}$.  We believe that SAR-Net plays an effective balance role in improving both classification and localization, which accords with the principle of SAR-Net.
Using the strong model Mask R-CNN with FPN+DCN, SAR-Net can also boost baseline by 2.7\% at AP$^{bb}_{50}$ on ChestX-Det. It demonstrates that our modules are robust enough.
Improvements on dataset DR-Private also demonstrates our effectiveness. It also means SAR-Net generalizes well on different data distributions.
It is noteworthy that the results on DR-private are slightly lower than the results on ChestX-Det. The main reason leading to different performance on two datasets is different data distribution of the two datasets. ChestX-Det from ChestX-14 has more bedside samples and DR-private has more regular ones. Abnormalities in bedside samples are often severer and salient.

We also report the performance of instance segmentation on both datasets. As shown in Table \ref{tab:instance seg}, our relation modules can also achieve considerable performance gain in the instance segmentation task.

\begin{table}
	\centering
	\caption{Main instance segmentation results on ChestX-Det and DR-Private.}
	\resizebox{86mm}{28mm}{
		\LARGE 
		\begin{tabular}{c|c|c|ccc}
			\hline
			\%    & Model & Method & \multicolumn{1}{c|}{AP$^{mask}$} & \multicolumn{1}{c|}{AP$^{mask}_{50}$} & AP$^{mask}_{75}$ \bigstrut\\
			\hline
			\hline
			\multirow{6}[6]{*}{\rotatebox{90}{ChestX-Det}} & \multirow{2}[2]{*}{Mask R-CNN(C4)} & Baseline & 13.8  & 32.6  & 9.9 \bigstrut[t]\\
			&       & SAR-Net & \textbf{15.3} & \textbf{35.4} & \textbf{12.9} \bigstrut[b]\\
			\cline{2-6}          & \multirow{2}[2]{*}{Mask R-CNN(FPN)} & Baseline & 16.2  & 36.6  & 13.4 \bigstrut[t]\\
			&       & SAR-Net & \textbf{17.0} & \textbf{38.6} & \textbf{14.1} \bigstrut[b]\\
			\cline{2-6}          & \multirow{2}[2]{*}{Mask R-CNN(FPN+DCN)} & Baseline & 16.2  & 38.3  & 12.3 \bigstrut[t]\\
			&       & SAR-Net & \textbf{17.9} & \textbf{40.6} & \textbf{14.9} \bigstrut[b]\\
			\hline
			\multirow{6}[6]{*}{\rotatebox{90}{DR-Private}} & \multirow{2}[2]{*}{Mask R-CNN(C4)} & Baseline &  13.8     &    30.2   & 12.4  \bigstrut[t]\\
			&       & SAR-Net &  \textbf{15.0}     &  \textbf{32.8}     & \textbf{13.8} \bigstrut[b]\\
			\cline{2-6}          & \multirow{2}[2]{*}{Mask R-CNN(FPN)} & Baseline &   16.9    & 37.3      & 13.8 \bigstrut[t]\\
			&       & SAR-Net &   \textbf{17.5}    & \textbf{38.0}      & \textbf{15.0}  \bigstrut[b]\\
			\cline{2-6}          & \multirow{2}[2]{*}{Mask R-CNN(FPN+DCN)} & Baseline & 17.5      &     37.1  & 15.4 \bigstrut[t]\\
			&       & SAR-Net &   \textbf{17.7}    & \textbf{37.9}      & \textbf{15.4} \bigstrut[b]\\
			\hline
	\end{tabular}}%
	\label{tab:instance seg}%
\end{table}%

\begin{table*}[t]
	\begin{center}
		\caption{Detection results Comparison at AP$^{bb}_{50}$ of each category on both datasets. The baseline model is Mask R-CNN with FPN+DCN.}
		\label{tab2}
		\resizebox{\textwidth}{!}{
			\renewcommand\arraystretch{1.1}
			\Large
			\begin{tabular}{c|c|c|c|c|c|c|c|c}
				\hline
				\%                                               & Model & Atelectasis & Calcification       & Cardiomegaly       & Consolidation  &  \makecell[c]{Diffusive\\ Nodule}          & Effusion   & Emphysema    \\ \hline 
				\multirow{6}{*}{\rotatebox{90}{ChestX-Det}}                       & Baseline  & $\bm{44.1}$          & 51.4         & 76.3          & 61.2          & 37.8              & 51.9          & 65.4          \\
				& SAR-Net  & 39.5   & $\bm{52.9}$ & $\bm{81.4}$  & $\bm{63.5}$  &$\bm{45.6}$      & $\bm{54.6}$  &$\bm{70.1}$ \\ \cline{2-9} 
				& Model &  \makecell[c]{Fibrosis}     & Fracture           & Mass         &  Nodule   & \makecell[c]{Pleural \\Thickening} & Pneumothorax  & $\bf{Mean}$  \\ \cline{2-9} 
				& Baseline  &$\bm{40.8}$            & 30.9 & 34.0          & $\bm{30.1}$         & 31.5              & 29.2         & 45.0          \\
				& SAR-Net  & 39.8   & $\bm{38.9}$       & $\bm{38.3}$  & 28.2 & $\bm{35.3}$      &$\bm{31.9}$ & $\bm{47.7}$  \\ \hline \hline
				\%                                               & Model & Atelectasis & Calcification       & Cardiomegaly       & Consolidation  &  \makecell[c]{Diffusive\\ Nodule}          & Effusion   & Emphysema    \\ \hline
				\multicolumn{1}{l|}{\multirow{6}{*}{\rotatebox{90}{DR-Private}}} & Baseline  &   11.6              &  $\bm{23.5}$            &     85.4          &  $\bm{55.5}$            &     43.5              &   46.1             &     39.2           \\ 
				\multicolumn{1}{l|}{}                            & SAR-Net  &      $\bm{12.0}$             &      23.4         &    $\bm{88.0}$            &    55.2           &    $\bm{45.0}$              &    $\bm{47.0}$           &     $\bm{48.2}$        \\ \cline{2-9} 
				\multicolumn{1}{l|}{}                            & Model &  \makecell[c]{Fibrosis}     & Fracture           & Mass         &  Nodule   & \makecell[c]{Pleural \\Thickening} & Pneumothorax   & $\bf{Mean}$  \\ \cline{2-9} 
				\multicolumn{1}{l|}{}                            & Baseline  &    $\bm{21.7}$               &      41.4          &    52.8            &       29.6        &      16.0              &     54.7           &     40.1           \\
				\multicolumn{1}{l|}{}                            & SAR-Net  &    19.9              &   $\bm{45.5}$           &  $\bm{48.5}$             &   $\bm{30.2}$           & $\bm{24.1}$                   &  $\bm{59.6}$              &     $\bm{42.0}$            \\ \hline
		\end{tabular}}
	\end{center}
\end{table*}

\begin{table*}[h!]
	\begin{center}
		\caption{Recall@0.1fp/image at AP$^{bb}_{25}$ of each category on both datasets.The baseline model is Mask R-CNN with FPN+DCN.}
		\label{tab3}
		\resizebox{\textwidth}{!}{
			\renewcommand\arraystretch{1.1}
			\Large
			\begin{tabular}{c|c|c|c|c|c|c|c|c}
				\hline                                               & Model & Atelectasis & Calcification       & Cardiomegaly       & Consolidation  & \makecell[c]{Diffusive\\ Nodule}           & Effusion   & Emphysema    \\ \hline  
				\multirow{6}{*}{\rotatebox{90}{ChestX-Det}}                       & Baseline  & $\bm{0.686}$           & 0.612         & 0.857          &$\bm{0.583}$           & 0.492              & 0.558          & 0.712          \\
				& SAR-Net  & 0.647   &$\bm{0.627}$  & $\bm{0.943}$& 0.576 & $\bm{0.603}$    &$\bm{0.616}$&$\bm{0.758}$\\ \cline{2-9} 
				& Model & \makecell[c]{Fibrosis}     & Fracture           & Mass         &  Nodule   & \makecell[c]{Pleural \\Thickening} & Pneumothorax  & $\bf{Mean}$  \\ \cline{2-9} 
				& Baseline  &$\bm{0.525}$            & 0.496 & 0.559          & 0.294          & 0.543              & $\bm{0.500}$      & 0.571          \\
				& SAR-Net  & 0.492   &$\bm{0.513}$       &$\bm{0.618}$ & $\bm{0.300}$ &$\bm{0.648}$    & 0.429 &$\bm{0.598}$  \\ \hline \hline                                               & Model & Atelectasis & Calcification       & Cardiomegaly       & Consolidation  & \makecell[c]{Diffusive\\ Nodule}          & Effusion   & Emphysema    \\ \hline 
				\multicolumn{1}{l|}{\multirow{6}{*}{\rotatebox{90}{DR-Private}}} & Baseline  &   0.341              &  0.394             &     0.921          &    0.608            &     0.633              &   $\bm{0.611}$            &     0.653           \\ 
				\multicolumn{1}{l|}{}                            & SAR-Net  &    $\bm{0.415}$            &   $\bm{0.437}$           & $\bm{0.943}$             &  $\bm{0.612}$             &  $\bm{0.653}$              &    0.575            &  $\bm{0.796}$          \\ \cline{2-9} 
				\multicolumn{1}{l|}{}                            & Model &  \makecell[c]{Fibrosis}      & Fracture           & Mass         &  Nodule   & \makecell[c]{Pleural \\Thickening}  & Pneumothorax   & $\bf{Mean}$ \\ \cline{2-9} 
				\multicolumn{1}{l|}{}                            & Baseline  & $\bm{0.382}$                &      0.422          & $\bm{0.651}$             &       0.379        &      0.329              &     0.731           &     0.543           \\
				\multicolumn{1}{l|}{}                            & SAR-Net  &    0.363             &   $\bm{0.452}$          &   0.605            & $\bm{0.407}$            &   $\bm{0.390}$                &  $\bm{0.731}$            &   $\bm{0.568}$           \\ \hline
		\end{tabular}}
	\end{center}
\end{table*}

	Table \ref{tab2} shows comparison results at AP$^{bb}_{50}$ of each category.
From our observation, pleural thickening, fracture, diffusive nodule, cardiomegaly and emphysema are categories with consistent and large improvements on both datasets. 
For representative categories of fracture and diffusive nodule, our method achieves AP$^{bb}_{50}$ of 38.9\% and 45.6\% on ChestX-Det, with a lead of 8.0\% and 7.4\%  over baseline respectively.
Fibrosis is the only category with decreasing accuracy in both datasets.
In our datasets, emphysema and fibrosis often exist simultaneously and have similar spatial distributions, both of which diffuse the whole lung fields.
Compared with fibrosis, the feature of emphysema is easier to distinguish, which cause that when fibrosis and emphysema overlap, the proposal boxes are easier to be predicted as emphysema.
When adding the relation modules, such phenomenon is more obvious.
The reason is that our SAR-Net has high sensitivity for the diseases which has more unique appearances, such as cardiomegaly and emphysema etc. 
Although the detection performance of fibrosis decreases slightly, the detection performance of emphysema obtains great improvement over both datasets.
For more practical usage, we also provide recall (sensitivity) at fixed false positives per image for each category in Table \ref{tab3}.
In particular, we set the rate of instance-level FP/image to be 0.1 at AP$^{bb}_{25}$.
From table \ref{tab3}, we can see that for most categories, our SAR-Net can obtain higher sensitivity than baseline.

\subsection{Performance on stronger backbones. }
We perform experiments with stronger backbones ResNet-101 and ResNeXt-101 \cite{resnext}. Results are shown in Table \ref{tab:backbone}. 

\begin{table}
	\centering
	\caption{Quantitative results via plugging our modules on different backbones. The baseline model is Mask R-CNN with FPN+DCN.}
	\label{tab:backbone}
	\resizebox{70mm}{18mm}{
		\normalsize 
		\begin{tabular}{c|c|ccc}
			\hline  \hline
			\%    & Backbone & AP$^{bb}_{25}$  & AP$^{bb}_{50}$  & AP$^{bb}_{75}$ \bigstrut\\
			\hline
			Baseline  & \multirow{2}[4]{*}{ResNet-50} & 54.8  & 45.0    & 16.1 \bigstrut\\
			\cline{1-1}    SAR-Net &       & \textbf{57.0} & \textbf{47.7} & \textbf{19.0} \bigstrut\\
			\hline
			Baseline  & \multirow{2}[4]{*}{ResNet-101} & 54.4  & 45.6  & 16.5 \bigstrut\\
			\cline{1-1}    SAR-Net &       & \textbf{57.9} & \textbf{48.5} & \textbf{19.7} \bigstrut\\
			\hline
			Baseline & \multirow{2}[4]{*}{ResNext-101 (32$\times$8d)} & 55.9  & 46.5  & 16.6 \bigstrut\\
			\cline{1-1}    SAR-Net &       & \textbf{58.1} & \textbf{48.9} & \textbf{20.0} \bigstrut\\
			\hline  \hline
	\end{tabular}}
\end{table}%

Our proposed SAR-Net consistently improves performance on all the experimented backbones. Specially, on the ResNet-101 backbone, we can obtain the gains of 3.5, 2.9 and 3.2 points on the AP$^{bb}_{25}$, AP$^{bb}_{50}$ and AP$^{bb}_{75}$ respectively. It demonstrates that our modules are robust enough and can be generalized to boost performance on multiple backbone architectures.

\subsection{Effectiveness of each relation module.}
In order to evaluate the effectiveness of each module, we conduct ablation studies on ChestX-Det from different perspectives (4.5-4.7). In all experiments, we use the same train, validation and test set. ResNet-50 is adopted as the backbone.
We remove the spatial relation module (SRM), the disease relation module (DRM) and the contextual relation module (CRM) from SAR-Net respectively. Ablation results are shown in Table \ref{tab4}.

	\begin{table}
	\caption{Ablation study on effect of each relation module. The baseline model is Mask R-CNN with FPN+DCN.}
	\label{tab4}
	\begin{center}
		\resizebox{80mm}{16mm}{
			\renewcommand\arraystretch{1.2}
			\begin{tabular}{c|ccc|ccc}
				\hline \hline
				\multirow{2}{*}{\%} & \multicolumn{1}{c|}{\multirow{2}{*}{\begin{tabular}[c]{@{}c@{}}SRM\end{tabular}}} & \multicolumn{1}{c|}{\multirow{2}{*}{\begin{tabular}[c]{@{}c@{}}DRM\end{tabular}}} & \multirow{2}{*}{\begin{tabular}[c]{@{}c@{}}CRM\end{tabular}} & \multicolumn{3}{c}{ChestX-Det}                           \\ \cline{5-7} 
				& \multicolumn{1}{c|}{} & \multicolumn{1}{c|}{} & & \multicolumn{1}{l}{AP$^{bb}_{25}$} & AP$^{bb}_{50}$ & AP$^{bb}_{75}$ \\
				\hline
				$\bf{SAR-Net}$  &  \checkmark     & \checkmark     &    \checkmark      & $\bm{57.0}$ & $\bm{47.7}$ & $\bm{19.0}$ \\
				\hline
				\multirow{2}{*}{} 
				& & 	\checkmark 		& 	\checkmark 		 & $57.0^{-0.0}$ & $47.3^{-0.4}$ & $17.5^{-1.5}$ \\  
				& \checkmark &  & 	\checkmark		 & $55.3^{-1.7}$ & $47.0^{-0.7}$ & $18.4^{-0.6}$ \\
				& \checkmark & \checkmark &  & $55.4^{-1.6}$  & $46.6^{-1.1}$ & $17.7^{-1.3}$  \\
				\hline \hline
		\end{tabular}}
	\end{center}
\end{table}

\textbf{The effect of spatial relation.} 
Compared with SAR-Net, if we remove the spatial relation module, the performance is decreased by 0.0\%, 0.4\% and 1.5\% on AP$^{bb}_{25}$, AP$^{bb}_{50}$ and AP$^{bb}_{75}$ respectively. The maximum decline on AP$^{bb}_{75}$ validates the effectiveness of the spatial relation module on localization. Moreover, the zero decline on AP$^{bb}_{25}$ validates that the spatial relation module has little positive effect on classification, which corresponds to the principle of the spatial relation module.

\textbf{The effect of disease relation.}
Compared with SAR-Net, removing the disease relation module decreases performance by 1.7\%, 0.7\% and 0.6\% on AP$^{bb}_{25}$, AP$^{bb}_{50}$ and AP$^{bb}_{75}$ respectively. 
This indicates that the disease relation module enhances feature representation more for classification than localization. When adding the disease relation module, some disease labels are corrected according to the information propagation among diseases. 

\textbf{The effect of contextual relation.}
Compared with the spatial relation+disease relation, adding the extra contextual relation module can boost the performance by 1.6\%, 1.1\% and 1.3\% on AP$^{bb}_{25}$, AP$^{bb}_{50}$ and AP$^{bb}_{75}$ respectively. The contextual relation module is encoded by spatial information and feature information.  This encoding mechanism has effectiveness on both classification and localization, which is verified by the experimental results. Moreover, the proposed attention mechanism can obtain more gains.

\subsection{Performance of each relation module on different diseases. }
In this subsection, we evaluate the performance of each relation module on different diseases. 
Specially, inspired by \cite{Chen}, we group all diseases into three super-classes, namely, LUNG, PLEURA and MEDIASTINUM. 
Detailed information of the super-classes and their sub-classes can be found in Table \ref{tab:datasets}.
Performance evaluations are conducted based on the super-classes and results are shown in Table \ref{tab:supercls}.

\begin{table*}[!htbp]
	\begin{center}
		\caption{Performance of each relation module on different diseases. The complexity of each module is also provided. The baseline model is Mask R-CNN with FPN+DCN and the backbone is ResNet50.}
		\label{tab:supercls}%
		\resizebox{\textwidth}{!}{
		\begin{tabular}{c|ccc|ccc|ccc|c|c}
			\hline
			\hline
			Super-Class   & \multicolumn{3}{c|}{LUNG} & \multicolumn{3}{c|}{PLEURAL} & \multicolumn{3}{c|}{MEDIASTINUM} & \multirow{2}[4]{*}{Params} & \multirow{2}[4]{*}{FLOPs} \bigstrut\\
			\cline{1-10}    Model & AP$^{bb}_{25}$  & AP$^{bb}_{50}$  & AP$^{bb}_{75}$  & AP$^{bb}_{25}$  & AP$^{bb}_{50}$  & AP$^{bb}_{75}$  & AP$^{bb}_{25}$  & AP$^{bb}_{50}$  & AP$^{bb}_{75}$  &  &  \bigstrut\\
			\hline
			Baseline & 52.0    & 44.0  & 15.5  & 55.9  & 37.5  & 5.9   & 76.3  & 76.3  & 52.9  &   -  & - \bigstrut[t]\\
			+ DRM & 52.8  & 44.7  & 16.9  & \underline{57.5}  & 38.3  & 5.1   & 78.7  & 76.8  & 50.4  & + 0.57M & + 0.18G \\
			+ SRM & 51.8  & 44.6  & 17.5  & 55.3  & 39.7  & \underline{6.2}   & 81.5  & 79.2  & \underline{\textbf{54.3}} & + 0.09M & + 0.26G \\
			+ CRM & \underline{53.6}  & \underline{44.8}  & \underline{18.0}    & 55.3  & 38.2  & 4.4   & 81.8 & 81.4  & 50.8  & + 3.29M & + 2.32G \bigstrut[b]\\
			\hline
			SAR-Net & \textbf{53.6}  & \textbf{46.3}  & \textbf{18.8}  & \textbf{58.1}  & \textbf{40.5}  & \textbf{9.8}   & \textbf{83.6}  & \textbf{81.4}  & 48.8  & + 3.8M  & + 2.5G \bigstrut\\
			\hline
			\hline
		\end{tabular}}%
	\end{center}
\end{table*}%

The contextual relation module obtains maximum gains on AP$^{bb}_{25}$, AP$^{bb}_{50}$ and AP$^{bb}_{75}$ respectively for LUNG diseases, while simply adding the spatial or disease relation module can not obtain great improvement. 
This is easy to understand, the diseases of LUNG distribute in the lung fields but most of them do not appear on fixed locations.
For instance, the atelectasis may appears in either the upper lobe or in the lower lobe, given different causes. 
While the contextual relation module are able to models contextual relations between diseases and observations in the lung fields, which involves both spatial and appearance compatibility, and therefore is more effective for the diagnosis of lung diseases. 

The diseases of PLEURA have strong co-occurrence relations. For instance, the effusion and primary spontaneous pneumothorax (PSP) often cause pleural thickening. 
Moreover, those three diseases have specific spatial distribution and locate around the pleura. Therefore, it is anticipated that the disease module and the spatial module should contribute more to the performance improvement. The experimental results also validated our hypothesis, as the disease module obtains gains on AP$^{bb}_{25}$ \& AP$^{bb}_{50}$ and the spatial module obtain gains on AP$^{bb}_{50}$ \& AP$^{bb}_{75}$. This suggests that the disease module can boost the performance for classification while the spatial module are better at improving localization accuracy, as we have emphasized before.

Compared with other diseases, cardiomegaly (MEDIASTINUM) has obvious appearance and fixed location, which makes it the perfect target disease of our proposed relation modules. It can be observed that all three relation modules can obtain considerable gains on AP$^{bb}_{25}$ and AP$^{bb}_{50}$. And the spatial relation module can also obtain gains on AP$^{bb}_{75}$, a metric that requires precise localization.

Different relation modules have different emphasis (classification or localization) for different diseases. 
In general, when adding all the relation modules, our model can achieve maximum performance.

\textbf{Model complexity}. We also report the complexity of each module in Table \ref{tab:supercls}.
As shown in Table \ref{tab:supercls}, using our modules can obtain higher performance with only a slight increase in parameters and floating point operations (FLOPs).
Note that some parameters are shared in different modules, such as the coordinate parameters of the anatomical structure.
Hence, the extra parameters of SAR-Net are less than the sum of extra parameters of all three modules. The same is true for FLOPs.

\subsection{Qualitative results. }
\begin{figure*}[t]
	\begin{center}
		\includegraphics[scale=0.38]{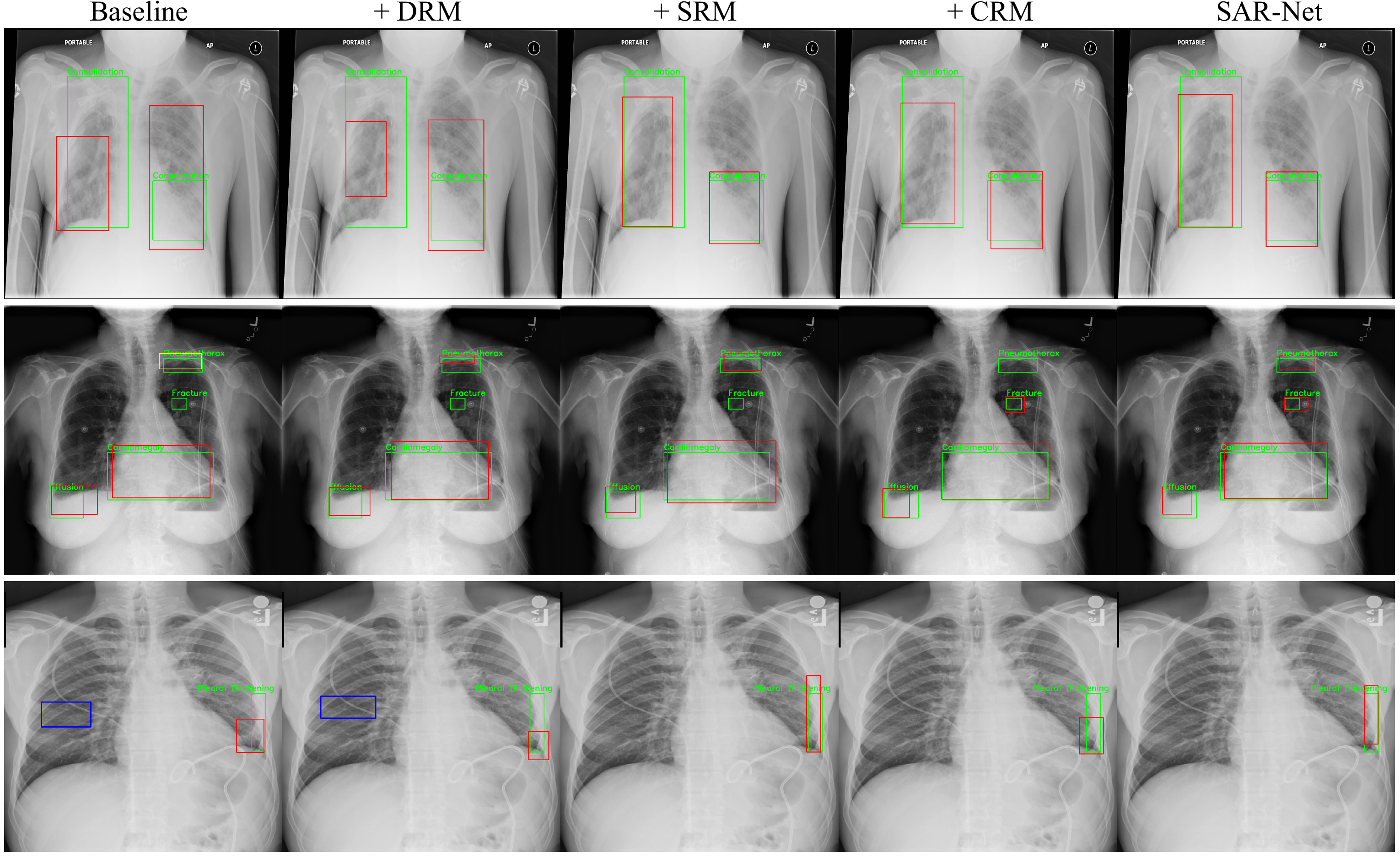}
	\end{center}
	\caption{Qualitative results on ChestX-Det. For fair comparison, the confidence threshold is chosen at FP/image=0.1.
		Green, yellow, red and blue boxes stand for ground-truths, false negatives, true positives and false positives respectively.}
	\label{fig:quality}
\end{figure*}

Figure~\ref{fig:quality} shows qualitative results of all the ablation methods. 
As shown in the first row, the disease relation module has little positive effect on localization.
However, the spatial relation module and contextual module can effectively improve the localization performance of the detected diseases. 
This is because all the above two relation modules are better at exploiting the spatial information. 

Pneumothorax and effusion have strong co-occurrence relations and specific spatial distribution and thus can benefit from the disease and spatial relation module. 
As shown in the second row, the disease relation module can correct the wrong disease labels according to the information propagation among diseases. On the other hand, the spatial relation module also can achieve the same effect with the help of the spatial information propagation. 
As have been shown on the quantitative results, the contextual relation module has high sensitivity on diagnosing the abnormalities of LUNG, but has low sensitivity on diagnosing the abnormalities of PLEURA. On this sample, the fracture is detected and pneumothorax is overlooked.

Pleural thickening develops when scar tissue thickens the delicate membrane lining the lungs (the pleura) and will not appear in the lung fields. 
As shown in the third row, the relation modules (SRM, CRM) which contain spatial information can eliminate this kind of false positives effectively. 
In addition, since the contextual relation module is not sensitive for PLEURA, it can not improve the localization performance of pleural thickening compared with the spatial relation module.

	\begin{floatingfigure}[r]{0.55\textwidth}
	\begin{center}
		\includegraphics[width=0.55\textwidth]{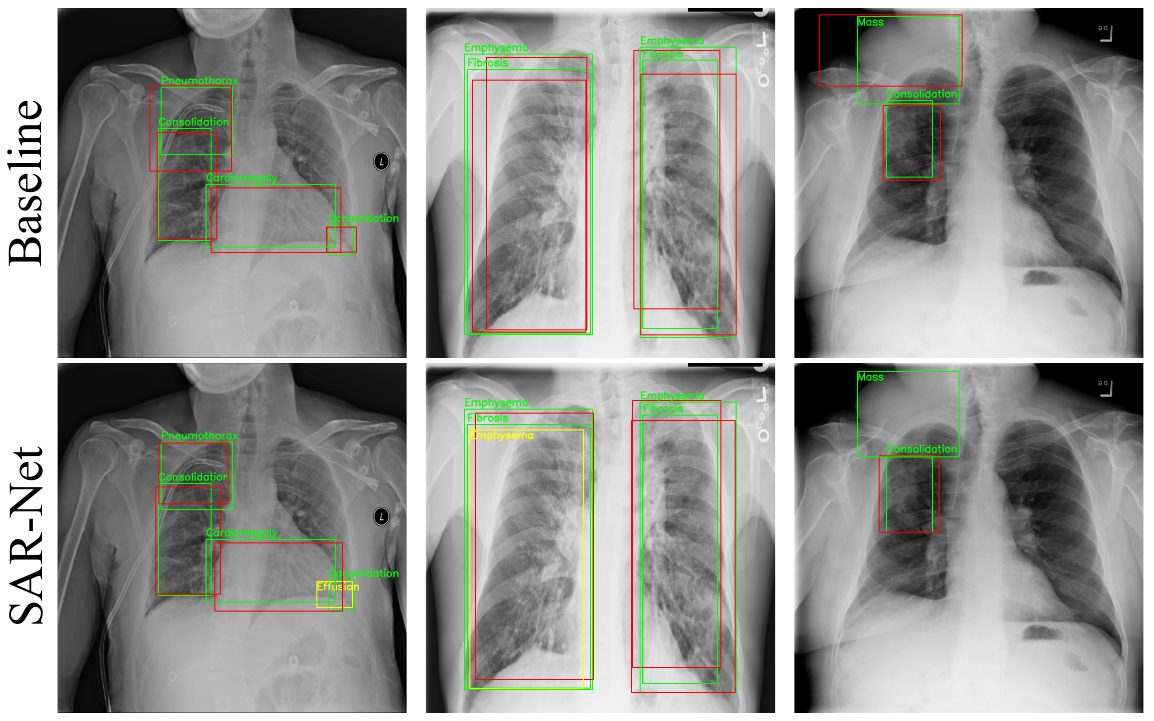}
	\end{center}
	\caption{
		Some failure cases on ChestX-Det. Green, yellow and red boxes stand for ground-truths, false negatives and true positives respectively.}
	\label{fig:fail_case}
\end{floatingfigure}

Moreover, we also show some failure cases in Figure \ref{fig:fail_case}. As mentioned before, effusion has strong co-occurrence relations with pneumothorax and often appears at costophrenic angle. As shown in the first case, when the relations modules are used, the above factors cause the consolidation box to be predicted as an effusion box. 
As mentioned in the subsection 4.3, when fibrosis and emphysema boxes overlap, the proposal boxes are easier to be predicted as emphysema. As shown in the second case, when the relations modules are used, the proposal box is predicted incorrectly and fibrosis is overlooked. In actual post-processing, this prediction box would be deleted by NMS. We keep it in Figure \ref{fig:fail_case} for illustration purpose. 
There exist a lymphoid mass around the neck in the third case. In our dataset, mass often appears in the lung fields, which causes the lymphoid mass to be overlooked when the relation modules are present. Note that such failure cases only occur occasionally.

Extensive experiment results indicate that our proposed method is effective and has a great potential for clinical application. 
In Figure \ref{fig:seg}, we also show some instance segmentation results of SAR-Net on ChestX-Det for close inspection. Only the contours are shown to retain the appearance of the detected diseases.

\section{Conclusion}
In conclusion, we present a structure-aware relation network (SAR-Net) on chest X-Ray detection and instance segmentation. The SAR-Net consists of three modules modeling three types of relations: 1. spatial relation between diseases and anatomical structure. 2. contextual relation between diseases and lung fields. 3. categorical relation within diseases. The proposed modules can be embedded into general object detection frameworks and bring significant improvements. Also, we present ChestX-Det, a subset of NIH ChestX-14 with box annotations of 13 categories of diseases. We believe the new dataset is a valuable benchmark for evaluation on disease detection in chest X-Rays.

\clearpage
\appendix

\bibliographystyle{splncs04}
\bibliography{reference}
\end{document}